\documentclass[usenatbib]{mn2e}

\usepackage{subfigure}
\usepackage{rotating}
\usepackage{ae,aecompl}
\usepackage{tikz}

\usepackage{graphicx}	
\usepackage{amsmath}	
\usepackage{amssymb}	

\newcommand{\AbellNiOnZe}{Abell~2261-910}
\newcommand{\AbellNiOnOn}{Abell~2261-911}

\newcommand{\AbellOnFoSiSe}{Abell~2261-1467}
\newcommand{\AbellOnFoSiEi}{Abell~2261-1468}
\newcommand{\AbellNiZeZeZe}{Abell~2261-9000}
\newcommand{\AbellOnFoZeZeZe}{Abell~2261-14000}
\newcommand{\AbellOnThSiSi}{Abell~2261-1366}

\newcommand{\AbellStacked}{Abell~2261-stacked}
\newcommand{\AbellNiZeZeZeMagnification}{4.3^{+1.6}_{-1.7}}
\newcommand{\AbellOnFoZeZeZeMagnification}{3.7^{+1.3}_{-1.3}}
\newcommand{\AbellOnThSiSiMagnification}{5.5^{+2.4}_{-2.3}}
\newcommand{\OII}{[O\,\textsc{ii}]}
\newcommand{\NIII}{[N\,\textsc{iii}]}

\newcommand\TnNiUpx{1.05}
\newcommand\TnNiUpy{1.3}
\newcommand\TnNiLox{1.1}
\newcommand\TnNiLoy{0.85}
\newcommand\TnFoUpx{0.90}
\newcommand\TnFoUpy{1.325}
\newcommand\TnFoLox{1.15}
\newcommand\TnFoLoy{0.85}
\newcommand\TnThUpx{1.40}
\newcommand\TnThUpy{1.35}
\newcommand\TnThLox{0.95}
\newcommand\TnThLoy{0.95}
\newcommand\TnRad{0.23}

\newcommand{\SFT}{SD}

\title[Multiply Lensed LAE in A2261]{A Multiply-Imaged $z \sim 6.3$ Lyman Alpha Emitter candidate behind Abell 2261}

\author[Rydberg et al.]{Claes-Erik Rydberg$^{1, 2}$\thanks{E-mail: mail@utte.nu},
Adi Zitrin$^{3, 4, 5}$,
Erik Zackrisson$^{6, 2}$,
Jens Melinder$^{2}$,
\newauthor
Daniel J. Whalen$^{7, 1}$,
Ralf S. Klessen$^{1, 8}$,
Juan Gonzalez$^{2}$,
G$\ddot{\mbox{o}}$ran $\ddot{\mbox{O}}$stlin$^{2}$,
\newauthor
and Daniela Carollo$^{9, 10}$
\\
$^{1}$Universit$\ddot{\mbox{a}}$t Heidelberg, Zentrum f$\ddot{\mbox{u}}$r Astronomie, Institut f$\ddot{\mbox{u}}$r Theoretische Astrophysik, Albert-Ueberly-Str. 2, 69120 Heidelberg, Germany\\
$^{2}$The Oskar Klein Center, Department of Astronomy, AlbaNova, Stockholm University, SE-106 91 Stockholm, Sweden\\
$^{3}$Cahill Center for Astronomy and Astrophysics, California Institute of Technology, MS 249-17, Pasadena, CA 91125, USA\\
$^{4}$Hubble Fellow\\
$^{5}$Physics Department, Ben-Gurion University of the Negev, P.O. Box 653, Be'er-Sheva 84105, Israel\\
$^{6}$Department of Physics and Astronomy, Uppsala University, Box 515, SE-751 20 Uppsala, Sweden\\
$^{7}$Institute of Cosmology and Gravitation, University of Portsmouth, Dennis Sciama Building, Portsmouth PO1 3FX, UK\\
$^{8}$Universit$\ddot{\mbox{a}}$t Heidelberg, Interdisziplin$\ddot{\mbox{a}}$res Zentrum f$\ddot{\mbox{u}}$r Wissenschaftliches Rechnen (IWR), Germany\\
$^{9}$Research School of Astronomy and Astrophysics, The Australian National University, Canberra, ACT 2611, Australia\\
$^{10}$INAF- Osservatorio Astronomico di Torino - Strada Osservatorio 20, Pino Torinese, 10020, Italy
}

\date{Accepted XXX. Received YYY; in original form ZZZ}

\pubyear{2016}

\begin{document}
\label{firstpage}
\pagerange{\pageref{firstpage}--\pageref{lastpage}}
\maketitle

\begin{abstract}
While the Lyman-$\alpha$ ($\mathrm{Ly}\alpha$) emission line serves as an important tool in the study of galaxies at $z\lesssim 6$, finding Ly$\alpha$ emitters (LAE) at significantly higher redshifts has been more challenging, probably because of the increasing neutrality of the intergalactic medium above $z\sim6$. Galaxies with extremely high rest-frame Ly$\alpha$ equivalent widths, EW(Ly$\alpha$)~$\gtrsim 150$~\AA{}, at $z>6$ are good candidates for Ly$\alpha$ follow-up observations, and can stand out in multiband imaging surveys because of their unusual colors. We have conducted a photometric search for such objects in the Cluster Lensing And Supernova survey with Hubble (CLASH), and report here the identification of three likely gravitationally-lensed images of a single LAE candidate at $z\sim6.3$, behind the galaxy cluster Abell~2261 ($z = 0.225$). In the process, we also measured with Keck/MOSFIRE the first spectroscopic redshift of a multiply-imaged galaxy behind Abell~2261, at $z = 3.337$. This allows us to calibrate the lensing model, which in turn is used to study the properties of the candidate LAE. Population~III galaxy spectral energy distribution (SED) model fits to the CLASH broadband photometry of the possible LAE provide a slightly better fit than Population~I/II models. The best fitted model suggests intrinsic EW(Ly$\alpha$) $\approx 160$~\AA{} after absorption in the interstellar and intergalactic medium. Future spectroscopic observations will examine this prediction as well as shed more light on the morphology of this object, which indicates it may be a merger of two smaller galaxies. 

\end{abstract}




\section{Introduction}
\label{intro}

Cluster lensing is a powerful tool for detecting and studying astronomical objects in the distant Universe \citep[see][for a recent review]{2011A&ARv..19...47K} and has the potential to bring otherwise undetectable objects within reach of existing or upcoming telescopes \citep[e.g.][]{2001ApJ...560L.119E, 2004ApJ...607..697K, 2010ApJ...717..257Z, 2012MNRAS.427.2212Z, 2012MNRAS.424L..54V, 2013MNRAS.429.3658R, 2013arXiv1312.6330W, 2014ApJ...793L..12Z, 2014ApJ...792...76B, 2014ApJ...795..126B, 2015ApJ...800...18A}.

\clearpage

For galaxies at redshifts $z\gtrsim 5$, the Ly$\alpha$ line is one of the few emission lines currently available for spectroscopic confirmation of photometric candidates \citep[although other alternatives exist, see for example][]{2014MNRAS.445.3200S, 2014ApJ...780L..18I, 2015ApJ...805L...7Z, 2016Sci...352.1559I, 2016ApJ...829L..11P}. However, LAEs are rare in the reionization era \citep[e.g.][]{2011ApJ...730....8H}, possibly because of the increasingly-neutral intergalactic medium at this epoch \citep[for a recent review, see][]{2014PASA...31...40D}. Until recently, no secure Ly$\alpha$ detections existed above $z \sim 7.5$ \citep{2012ApJ...744..179S, 2013Natur.502..524F}, but lately more than a handful of spectroscopic detections have been reported at $z \gtrsim 7.5$ \citep[e.g.][]{2012ApJ...744...83O, 2013Natur.502..524F, 2015Natur.519..327W, 2015ApJ...804L..30O, 2015ApJ...810L..12Z, 2016ApJ...826..113S, 2016arXiv160303222K}, with the highest-redshift LAE currently known at $z=8.68$ \citep{2015ApJ...810L..12Z} and the earliest Lyman break galaxy, whose break was verified further by HST grism, at $z=11.1$ \citep{2016ApJ...819..129O}.

At somewhat lower redshifts, a fraction of LAEs exhibit very high Ly$\alpha$ equivalent widths \citep[extreme LAEs, with EW(Ly$\alpha$) $\gtrsim 150$ \AA{}; e.g.][]{2002ApJ...565L..71M, 2011ApJ...734..119K}. They are intriguing because they could in principle host populations of metal-free (Pop~III) stars \citep[e.g.][]{2002A&A...382...28S, 2010A&A...523A..64R}. Other explanations for these extreme LAEs include gas cooling \citep{2009ApJ...690...82D}, directional EW(Ly$\alpha$) boosting \citep{2014MNRAS.444.1095G}, accreting black holes \citep{2001ApJ...556...87H} and stochastic sampling of the stellar initial mass function \citep{2013MNRAS.428.2163F}. \citet{2012ApJ...761...85K} present a LAE with an observed EW(Ly$\alpha$) of 436$^{+422}_{-149}$~\AA{} at $z=6.538$. The Ly$\alpha$ line is spectroscopically measured and the continuum is detected in their deep $z$'-band image. In \citet{2015ApJ...808..139S} two objects, CR7 and MASOSA, were presented, both with $\mathrm{EW(Ly}\alpha) \gtrsim$ 200 \AA{} at $z_{\mathrm{spec}}=6.604$ and $z_{\mathrm{spec}}=6.541$, respectively. To our knowledge, the \citet{2012ApJ...761...85K} object and CR7 are the first extreme LAEs detected above $z\sim 6.5$. \citet{2015ApJ...808..139S} conclude that CR7 is best explained by one population of mostly Pop~III stars and two populations of Pop II/I stars (metallicity $\mathrm{Z} > 0$). However, \citet{2015MNRAS.453.2465P} as well as \citet{2016MNRAS.460.4003A} and \citet{2016MNRAS.462.2184H} argue that the observations are better explained by a direct collapse black hole accreting primordial gas, a possibility also briefly discussed in \citet{2015ApJ...808..139S}. \citet{2016arXiv160900727B} object to both interpretations, claiming that deeper observations in Spitzer Space Telescope (Spitzer) 3.6 and 4.5 micron bands indicate strong [O\,\textsc{iii}] emission. They suggest two alternative interpretations, either a Type II AGN or a low-metallicity starburst.

A plausible triple galaxy merger at $z = 6.595$ dubbed 'Himiko' was presented in \citet{2009ApJ...696.1164O, 2013ApJ...778..102O}. Alhough not an extreme LAE, it has strong Ly$\alpha$ emission with EW(Ly$\alpha$) = 78$^{+8}_{-6}$~\AA{}. Its Ly$\alpha$ halo was detected with a narrow band filter and appears as a 'blob' covering all three galaxies. The authors use ALMA to observe the [C\,\textsc{ii}] line as it is a tracer of star forming regions and could potentially reveal kinematics of the merger. However, no line was observed that would suggest Himiko to be a metal poor object. \citet{2016ApJ...823L..14H} present a triply imaged galaxy behind the galaxy cluster MACS 2129. Like Himiko, it has similarly strong Ly$\alpha$ emission with EW(Ly$\alpha$) = 74$\pm 15$~\AA{} and a spectroscopic redshift of 6.85.

In \citet{2011MNRAS.418L.104Z}, we argue that extreme LAEs could in principle be identified from HST broadband data even at redshifts up to $z \sim 8-9$ because of their unusual colors. The key point is that the relevant emission lines are so prominent that the broadband to which the line is redshifted would appear significantly brighter because of the additional flux from the line. Since we know (in part from SED models) which lines are typically stronger in star-forming young galaxies, we can anticipate the effect of brightening on the broadband photometry at a given redshift. This selection technique is similar in nature to that recently adopted by \citet{2014ApJ...784...58S,2015ApJ...801..122S} and \citet{2016ApJ...823..143R} to successfully search for $z \sim 6-8$ galaxies with strong rest-frame optical emission lines. For example, all 4 objects at $z>7$ predicted by \citet{2016ApJ...823..143R} to be prominent emission line galaxies from their unusual broadband colors were later found to exhibit Ly$\alpha$ in follow-up observations \citep{2015ApJ...804L..30O, 2015MNRAS.454.1393S, 2015ApJ...810L..12Z, 2016ApJ...823..143R}.

The CLASH survey \citep{2012ApJS..199...25P} has now produced an extensive broadband data set for the identification of lensed galaxy candidates at high redshifts \citep[e.g.][]{2012Natur.489..406Z, 2013ApJ...762...32C, 2014ApJ...792...76B, 2014ApJ...783L..12V}. In a companion paper \citep{2015ApJ...804...13R} we presented a search for Pop~III LAE candidates at $z\gtrsim 6$ in CLASH, which resulted in two singly-imaged candidates with best-fitting redshifts at $z\approx 8$. These candidates -- if confirmed spectroscopically -- might like the aforementioned Roberts-Borsani objects be the highest-redshift extreme LAEs detected so far.

We have now discovered another interesting object from CLASH: a multiply-imaged, high-magnification extreme LAE candidate with an estimated photometric redshift of $z \sim 6.3$. We have detected three images in CLASH Abell~2261 data that show binary substructure that, with SED fitting with different models, satisfy our criteria for an extreme LAE. We use a lens model for Abell~2261 which we revise here following our spectroscopic measurement of the first multiply-imaged galaxy behind this cluster. Using this in combination with the photometric redshifts of our three images we conclude it is likely that the three images are the same object. Here, we present this candidate, its predicted extreme LAE properties, and our revised lens model. Future observations targeting Ly$\alpha$ in these objects will soon be able to test these predictions. Substructure is observed in the object which could be investigated by additional surveys of the [C\,\textsc{ii}] line \citep{2016MNRAS.462L...6K}, where a splitting of the line could indicate a recent merger.

Throughout this paper, we assume a $\Lambda$CDM Universe with cosmological parameters H$_0 = 67.3$, $\Omega_\mathrm{M}=0.308$, and $\Omega_{\Lambda}=0.692$ based on Planck, WP (WMAP polarization), highL (high resolution CMB), and baryon acoustic oscillation data \citep{2014A&A...571A..16P}. In Section~\ref{sec:observations} we review the available CLASH, Spitzer satellite, and spectroscopic data sets and in Section~\ref{sec:models} we discuss the stellar population models applied in our study.  The gravitational lens model we construct is described in Section~\ref{sec:gravitationallensing}. The analysis of the LAE is presented in Section~\ref{sec:results}, and we conclude in Section~\ref{sec:summaryconclusion}.

\section{Observations}
\label{sec:observations}

\subsection{Imaging}
\label{sec:observationaldata}

The CLASH survey \citep{2012ApJS..199...25P} has provided deep Hubble imaging data of 25 galaxy clusters, including Abell~2261. The observations of Abell~2261 ($z = 0.225$) cover a broad wavelength range (2,000~-~17,000~\AA{}) in 16 filters from the UV to the near-infrared. A photometric catalog for Abell~2261, generated using SExtractor \citep{1996A&AS..117..393B}, was made publicly available as a high-end CLASH product \citep{2012ApJS..199...25P}. The catalog contains 2,127 potential objects, with data in the 16 filters. Note that the ID numbers for the LAE candidates we use throughout, originated from this catalog.

Abell~2261 has also been imaged with the Spitzer satellite. In our study we use the publicly available\footnote{http://sha.ipac.caltech.edu/applications/Spitzer/SHA/} images from the Infra Red Array Camera \citep[IRAC;][]{2004ApJS..154...10F}  3.6, 4.5, 5.8, and 8.0~micrometer filters.


\subsection{Spectroscopy}
\label{sec:spectroscopicdata}

Spectroscopic observations of Abell~2261 were carried out with the Multi-Object Spectrometer For Infra-Red Exploration \citep[MOSFIRE; ][]{2012SPIE.8446E..0JM} on the Keck 1 telescope, as part of author A. Zitrin's search for UV metal lines in $z\sim7-8$ lensed objects \citep[described in][]{2015ApJ...805L...7Z}. Abell~2261 was observed on 2014, September 16, for 2.2 hours in the H band consisting of sets of 120s exposures. An AB dither pattern of $\pm1.25\arcsec$ along the slit was used and the typical seeing was $\sim0.6\arcsec$. All spectroscopic data were reduced using the public MOSFIRE pipeline. For each reduced slit (biased, flat-fielded and combined), the 1D spectrum was then extracted using an 11~pixel boxcar  ($\sim1\arcsec$) centered on the target. The 1$\sigma$ error was extracted using the same procedure, in quadrature.

\begin{figure*}
    \begin{center}
        \includegraphics[width=160mm, trim= 0cm 7cm 0cm 4cm,clip]{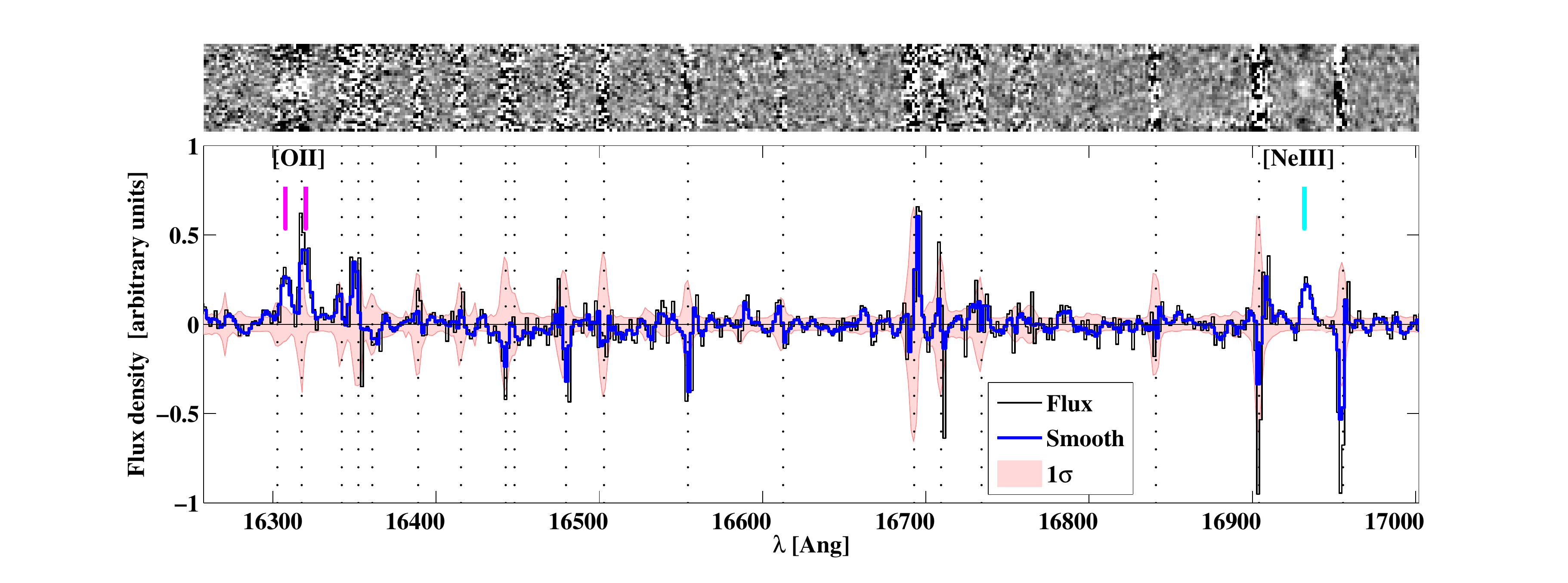}
        \caption{
H-band spectrum of a multiply imaged source at $z=3.377$ lensed by Abell~2261, taken with MOSFIRE on the Keck 1 telescope (see Section~\ref{sec:spectroscopicdata}). The upper panel shows the 2D spectrum, whose 1D boxcar extraction is shown in the bottom panel along with a slightly smoothed version and the $1\sigma$ error. We show here only the central part of the H-band showing the detection of the \OII{} (3727, 3729~\AA{}) doublet, and \NIII{} (3869~\AA{}) line, at $z=3.377$. This redshift measurement constitutes the first measurement of a multiply-imaged source in Abell~2261 and allows us to anchor the redshift calibration of its lens model. Using the revised model we then examine the possibility that the 3 objects investigated in this work are images of the same $z\sim6.3$ background source (see Section~\ref{sec:counterimagecomparison}).
}
        \label{fig:spectroscopicredshift}
    \end{center}
\end{figure*}

The H-band (14,500--17,770~\AA{}) mask included a slit placed on multiple image 4a at (17:22:28.56, +32:08:07.92), using the ID given in \citet{2012ApJ...757...22C}. The image was identified initially using the method of \citet{2009MNRAS.396.1985Z}, which is the one used here for lens modeling as well. \citet{2012ApJ...757...22C} obtained a photometric redshift of $3.48\pm0.03$ for this galaxy, and the lens models presented therein similarly predict $z\sim3.3$. Our spectroscopic observations, shown in Figure~\ref{fig:spectroscopicredshift}, indicate a redshift of $z=3.377$ based on the [O\,\textsc{ii}] (3727, 3729~\AA{}) doublet, and [N\,\textsc{iii}] (3869~\AA{}) line, in good agreement with the photometric and initial lens model prediction. This redshift measurement is important, and will allow us to recalibrate (for the first time) the lens model of Abell~2261. The H-band observations also covered one of the objects presented in this paper\footnote{Later dubbed \AbellOnThSiSi{}.}, although no lines were detected.

We have also examined J-band (11,500-13,520~\AA{}) observations covering two of the objects presented\footnote{\AbellNiZeZeZe{} and \AbellOnFoZeZeZe{}.}. The J-band observations were carried out on June 10 and 11, 2015, for a total of 2.9 hours. No significant lines were detected.

Detections of metal-lines in the H or J band could have ruled out the possibility of the galaxy being metal-free. The J-band observations also weakens the plausibility of the object being a low-redshift galaxy, see Section~\ref{sec:results}.

\bigskip

\section{Spectral Energy Distribution Models}
\label{sec:models}

To derive various physical properties and the redshifts of galaxies, model spectra are usually fitted to the photometric data. The models can be either synthetic spectra, empirical spectra, or a mix of the two. We consider here four different grids of spectra to fit to the observed data, which we now describe.

\subsection{Yggdrasil}
\label{sec:modelsyggdrasil}

The Yggdrasil population synthesis model is described in detail in \citet{2011ApJ...740...13Z} so only a brief summary is given here. As input for Yggdrasil we use Starburst99 \citep{1999ApJS..123....3L, 2005ApJ...621..695V} stellar component spectra for instantaneous bursts based on stellar evolutionary tracks from the Padova group \citep{2005ApJ...621..695V}. Nebular emission (in the form of emission lines and nebular continuum radiation) is added using the photoionization code \textsc{Cloudy} \citep{1998PASP..110..761F}. The Yggdrasil model grid adopted in this paper features the following parameters: 

\begin{description}

\item[{\bf Covering factor ($f_\mathrm{cov}$)}] This parameter regulates the contribution of nebular emission to the model spectrum and represents the fraction of the stellar component covered by photoionized gas \citep[see][for a more detailed description]{2013ApJ...777...39Z}. We consider $f_\mathrm{cov}=0.0$ (no nebula; direct starlight only), 0.5 (intermediate case), 1.0 (maximal contribution from ionized gas).
\item[{\bf Initial mass function (IMF)}] Three IMFs are considered: The standard Kroupa IMF \citep{2001MNRAS.322..231K}, a log-normal distribution with typical mass $M_{\mathrm{typ}} \sim 10~\mathrm{M}_{\odot}$ with $\sigma = 1.0$ for the underlying normal distribution, and a top-heavy distribution with $M_{\mathrm{typ}} \sim 100~\mathrm{M}_{\odot}$ with Salpeter slope in [50,~500]~M$_{\odot}$. The latter two distributions are only considered for Pop~III stars.

It is worth noting that in recent studies of the Milky Way halo stellar populations \citep[][ and references therein]{2014ApJ...788..180C}, alternative IMFs were considered to take into account the properties of the inner- and outer halo populations \citep{2007Natur.450.1020C, 2010ApJ...712..692C} and their building blocks. The Milky Way's halo comprises at least two diffuse stellar populations with different spatial distribution, kinematics, dynamics, and chemical composition. Such properties  suggest that inner- and outer-halo formed through distinct astrophysical mechanisms and their building blocks were substantially different: more-massive dwarf galaxies with large content of gas and sustained star formation for the inner-halo, lower-mass dwarf galaxies with low content of gas and truncated star formation for the outer-halo.

In a more recent investigation \citet{2014ApJ...788..180C} analyzed the distribution of two main classes of carbon enhanced metal poor stars \citep[CEMP-s and CEMP-no; ][]{2005ARA&A..43..531B} in the two halo components. They found that the outer-halo possesses a much larger fraction of CEMP-no stars, while the inner-halo exhibits a fraction of CEMP-s twice the fraction of CEMP-no. These two CEMP classes had different progenitors: intermediate-mass stars (1.3--3.5~M$_{\odot}$) for the CEMP-s, and massive progenitors (10--60~M$_{\odot}$) for the CEMP-no. Because of such distinct chemical signature \citet{2014ApJ...788..180C} argued that it is likely that the outer halo dwarf galaxies progenitors possessed a flatter IMF with a slope $\alpha = -1.5$. On the contrary the inner halo dwarf galaxies progenitors had a standard IMF with slope $\alpha = -2.35$ (assuming a Salpeter power law). 

The implemented IMFs in the present investigation spans a wide range of masses, however the quality of the fits doesn't show significant variations. Populations synthesis models that employ the alternative IMFs derived from Milky Way's studies could impact the mass estimate and the results of our fits, and will be considered in future papers.

\item[{\bf Starburst duration (\SFT{})}] Four values of \SFT{}s are considered: 0~Myr (i.e. all stars formed instantaneously), 10~Myr, 30~Myr, and 100~Myr. For Pop~III galaxies only 0~Myr is used since results for longer \SFT{}s are nearly degenerate for young galaxies.
\item[{\bf Age}] The age parameter has a range from 1~Myr to the age of the universe at the given redshift.
\item[{\bf Metallicity}] In this paper, we use Yggdrasil models with Z~=~0, 0.02~$\mathrm{Z}_{\odot}$, 0.2~$\mathrm{Z}_{\odot}$, 0.4~$\mathrm{Z}_{\odot}$, and 1.0~$\mathrm{Z}_{\odot}$. They are applied in two grids: one Pop~III galaxy grid (Z~=~0) and one grid with $\mathrm{Z} > 0$. The purpose is to compare the quality of fit to the two different grids to investigate wether the observations correspond to a Pop~III galaxy or a galaxy containing metals.
\item[{\bf Lyman-$\alpha$ escape fraction $f_{\mathrm{Ly}\alpha}$}]  This parameter varies from 0.0 to 0.5, and represents the combined escape fraction of Ly$\alpha$ photons from the interstellar medium (ISM) and the intergalactic medium (IGM). See Section~\ref{sec:lyaline}.

\end{description}

\subsection{Gissel}
\label{sec:gissel}

This is a grid of synthetic spectra from \citet{2003MNRAS.344.1000B}. Gissel contains models with different metallicities (but not zero metallicities, i.e. no Pop~III stars), ages and extinction. These models do not contain nebular emission. To account for dust attenuation we use three different attenuation curves \citep{2000ApJ...533..682C, 1984A&A...132..389P, 1979MNRAS.187P..73S}.

\subsection{CWW, Kinney}

The CWW, Kinney grid \citep{1980ApJS...43..393C, 1996ApJ...467...38K, 1999MNRAS.310..540A} builds partly on empirical spectra. The base is UV observations of nearby galaxies which are then extrapolated with the Gissel code. Starburst galaxies with emission lines are included. Similarly to Section~\ref{sec:gissel} we use three different attenuation curves \citep{2000ApJ...533..682C, 1984A&A...132..389P, 1979MNRAS.187P..73S} to simulate dust attenuation.

\subsection{The Lyman-$\alpha$ line}
\label{sec:lyaline}

The Ly$\alpha$ line is produced by recombination from the first excited state to the ground state in hydrogen. For Yggdrasil models with nebular emission (i.e. $f_{\mathrm{cov}} > 0$), significant intrinsic Ly$\alpha$ emission is produced. However, the Ly$\alpha$ line is highly resonant. When a Ly$\alpha$-photon encounters neutral hydrogen it is easily absorbed, and re-emitted in a random direction. This scattering process makes Ly$\alpha$ very susceptible to absorption by dust in the ISM because of its increased path length through the galaxy. In the IGM the scatter could reduce the observed Ly$\alpha$ radiation or render it undetectable. The parameter $f_{\mathrm{Ly}\alpha}$ is used to represent the fraction of Ly$\alpha$ photons escaping both the ISM and the IGM. Simulations imply an upper bound of 0.5 on this parameter at $z > 6$ \citep{2011MNRAS.414.2139D}. The equivalent widths (EW) presented in Table~\ref{tab:RT1} are rest-frame EWs compensated by $f_{\mathrm{Ly}\alpha}$, hence cosmological expansion has not been taken into account.

Shorter wavelength radiation that is redshifted to the Ly$\alpha$-wavelength ($\lambda_{\mathrm{Ly\alpha}}=1,216$~\AA{}) is also scattered or absorbed if it encounters neutral hydrogen. Since the universe is increasingly neutral at $z>6$, virtually all radiation below $\lambda_{\mathrm{Ly\alpha}}$ is absorbed. This Gunn-Peterson trough \citep{1965ApJ...142.1633G} is approximated here by setting the flux below $\lambda_{\mathrm{Ly\alpha}}$ to zero for all model spectra at $z>6$. At $z<6$, the situation is more complicated since the radiation encounters distinct clouds of neutral hydrogen. To simulate the absorption by these clouds we use the model by \citet{1995ApJ...441...18M}. The model provides the average absorption as a function of redshift and wavelength not only from Ly$\alpha$ absorption but also from absorption by higher-order Lyman lines.

\section{Lens model}
\label{sec:gravitationallensing}

Since the examined LAE images are lensed by a cluster, we use here a lens model to derive their magnification and examine whether the different images are related to the same LAE source. The spectroscopic measurement mentioned above (Section \ref{sec:spectroscopicdata}) provides the first redshift for a multiple-imaged galaxy in this cluster and thus allows us to calibrate and refine an existing lens model for this cluster published previously \citep{2015ApJ...801...44Z}. 

The lens model we use here is thus an updated version of the models \citep{2015ApJ...801...44Z} available via the Mikulski Archive for Space Telescopes (MAST) as a high-end CLASH science product \citep{2012ApJS..199...25P}. We use the light-traces-mass (LTM) approach by \cite{2005ApJ...621...53B} and \cite{2009MNRAS.396.1985Z, 2015ApJ...801...44Z}.

The basic LTM model is constructed from two mass components, supplemented with an external shear.  The first mass component corresponds to the contribution of cluster galaxies, and the second one to the dark matter (DM) contribution. We start by identifying cluster-member galaxies by following the red-sequence in a color magnitude diagram. Each galaxy is assigned with a power-law mass density distribution projected two dimensionally, scaled by its luminosity. The superposition map of the contributions from all cluster galaxies is then smoothed with a 2D Gaussian kernel, to obtain the smooth, light-tracing-mass DM component. The two components are then added with a relative weight, and supplemented by a 2-parameter external shear.  This basic model has 6 free parameters: the exponent of the galaxy power-law (which is free, but the same for all galaxies); the width of the Gaussian kernel, the relative galaxy-to-DM weight, the direction and amplitude of the external shear, and an overall normalization of the model. 

The best-fit parameter values are derived through a several-thousand long Monte-Carlo Markov Chain (MCMC), using $\chi^2$ criteria for the reproduced positions of multiple images. We use here the list of multiple images presented in \citep[][see also \citealt{2012ApJ...757...22C}]{2015ApJ...801...44Z}. Most systems lack spectroscopic redshift which means these are left to be freely optimized by the model around their photometric redshift value. To allow for further flexibility in the reproduction of images, note that we also allow here for the mass and ellipticity of the brightest cluster galaxy to be freely optimized in the minimization procedure.

We use a $\chi^2$ multiple-image positional error of 0.5$\arcsec$ for the minimization, and errors are extracted using 50 realizations of the MCMC model. Note that in \citet{2015ApJ...801...44Z} we quantified that the error maps with this positional uncertainty are likely underestimated, and we combine this positional uncertainty with the systematic uncertainty thus estimated by taking the square root of the squared sum. The systematic uncertainty level was found to be more representative of shifts caused by random structure along the line of sight \citep{2012MNRAS.420L..18H} and encompass better the differences between different lens modeling techniques.

As constraints we use only the 4 most secure systems listed in \citet{2012ApJ...757...22C} and \citet{2015ApJ...801...44Z}, but update and fix the redshifts of systems 4 and 5 to the redshift we measure here for image 4.a. We use this model to help us understand better if these are likely counter images of the same source, which were not used as constraints. The final model has an image reproduction rms of 0.60$\arcsec$. The prediction for the positions of the LAE images is discussed in Section~\ref{sec:results}.

\begin{figure*}
  \begin{center}
	  \subfigure[Abell~2261]{\label{fig:counterimageareas}\includegraphics[scale=0.32]{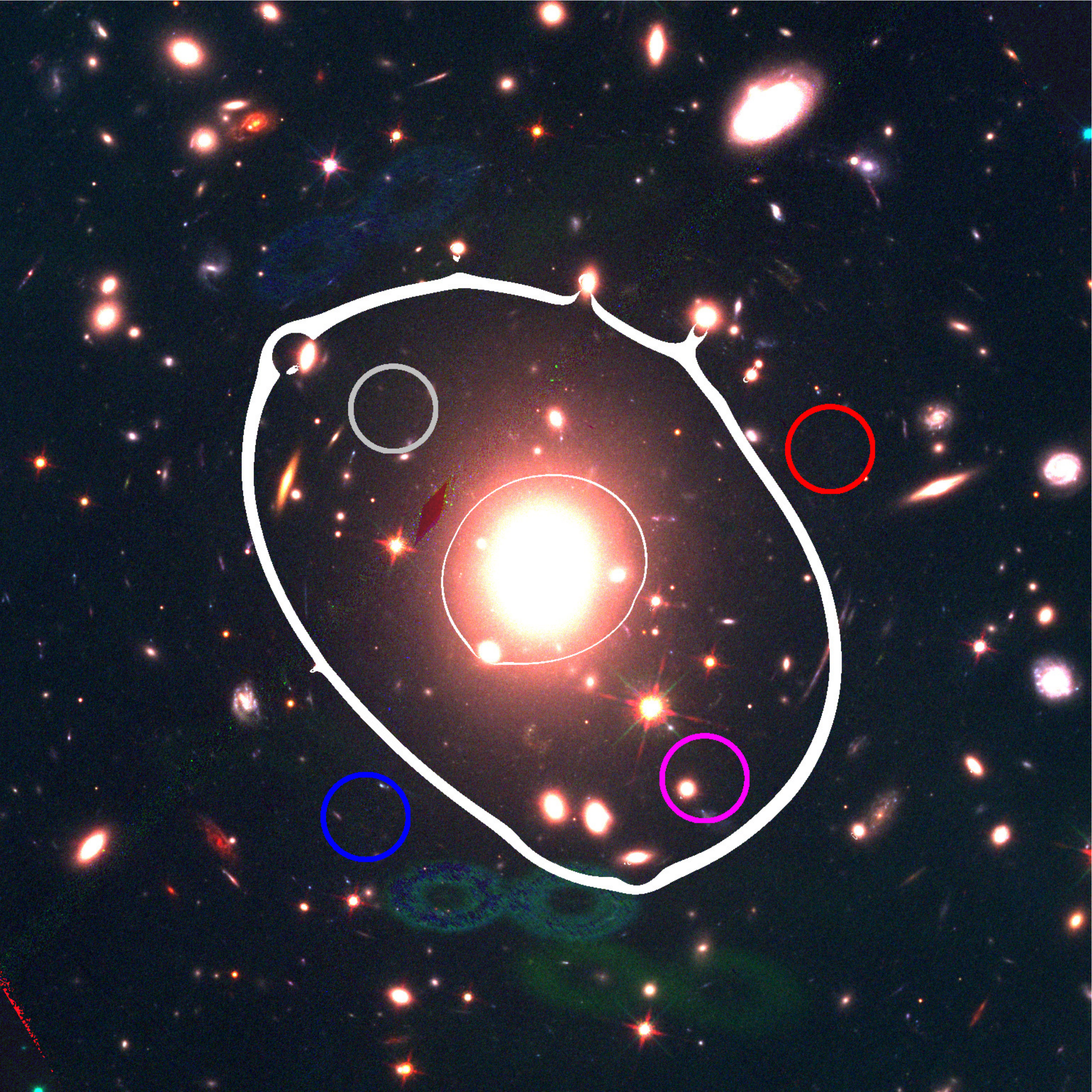}}
		\subfigure[Abell~2261, magnification map]{\label{fig:magnificationmap}\includegraphics[scale=0.3215]{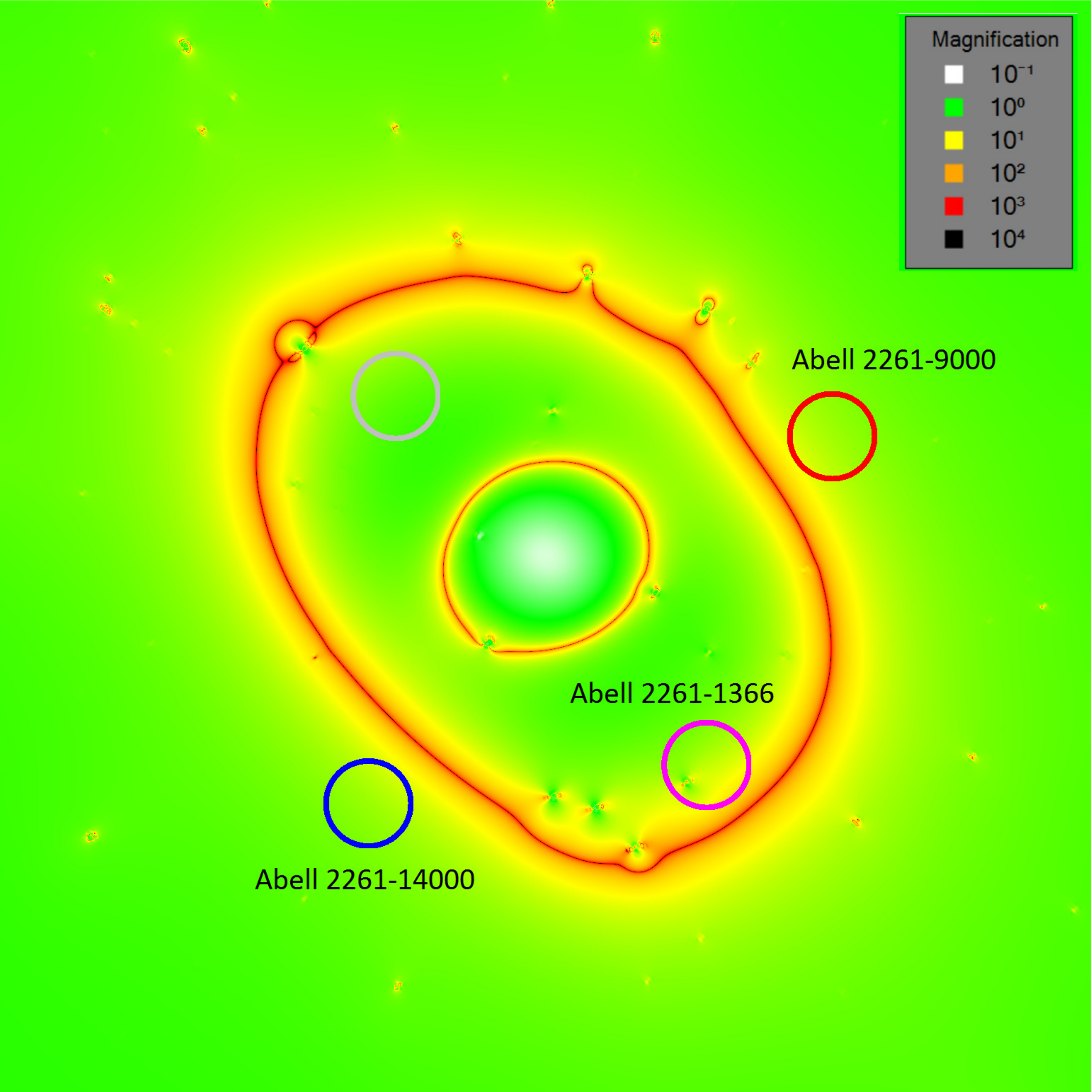}}
  \end{center}
  \caption{
$100\arcsec \times 100\arcsec$ overviews. The left panel is a multicolor image of Abell~2261 (F606W is represented by blue, F814W is represented by green, and F125W is represented by red) of Abell~2261. The critical line for $z_{\mathrm{s}}=6.3$ is marked in white. The right image is a magnification map for Abell~2261 \citep{2015ApJ...801...44Z}, color-coded to show the magnification for a source at $z_{\mathrm{s}}=6.3$. The originally identified object, \AbellNiOnZe{}, and a second nearby object, \AbellNiOnOn{}, at a similar photometric redshift are marked by a red circle. The regions where counter-images were predicted to be are marked in blue, purple, and gray. Each circle has a radius of $4\arcsec$ and is centered where the counter-image is predicted to be which, because of prediction errors, is not necessarily exactly on the counter-image candidate(s). In the blue area the two plausible counter-images \AbellOnFoSiSe{} and \AbellOnFoSiEi{} (at an angular distance of $1.7\arcsec$ and $0.9\arcsec$ from the counter-image's predicted position) were found to correspond in redshift estimate as well as morphology to the two objects in the red area. In the purple area the plausible counter-image \AbellOnThSiSi{} at an angular distance of $3.2\arcsec$ from the predicted position were identified. In the gray area no plausible counter-images were found.
}
  \label{fig:a2261overview}
\end{figure*}

\section{Results}
\label{sec:results}

Our LAE candidate was identified when searching for Pop~III galaxies in the CLASH data. The method, described in more detail in \citet{2015ApJ...804...13R}, uses $\chi^2$ as well as cross-validation \citep[see][]{Singh1981} to fit the observational data to the four different model grids (Section~\ref{sec:models}). The $\chi^2$-fitting was carried out with the publicly available \textsc{Le Phare} \citep{2006A&A...457..841I} code while cross-validation was done with a program we developed called Observational Data Scanner (ODS). The two methods produce very similar results. The CLASH dataset was then scanned for objects with good quality of fits to Pop~III model galaxies. The fits should also, preferably, be significantly better than galaxy models containing metals. The search also included the criterion that the objects must be sufficiently extended to rule out point sources.

In this search, \AbellNiOnZe{} \citep[previously published in ][]{2014ApJ...792...76B} was initially flagged as a potential Pop~III candidate, although its likely multiply lensed nature was not known at the time \citep{2014MmSAI..85..210R}. Although it was later found not to be a convincing Pop~III galaxy candidate because the quality of the fit for Yggdrasil $\mathrm{Z}>0$ models was considered to be comparable, its multiply lensed nature was revealed in the analysis. In the region containing \AbellNiOnZe{} a nearby object \AbellNiOnOn{} was identified at a similar redshift and may be part of the same system. Because of their proximity, \AbellNiOnZe{} and \AbellNiOnOn{} are treated as image substructures of just one object (hereafter \AbellNiZeZeZe{}, see Section~\ref{sec:abell9000}). Using the redshifts implied from Yggdrasil fits and the lensing model (Section~\ref{sec:gravitationallensing}), the expected positions of counter-images for \AbellNiZeZeZe{} were derived. The four positions within the Abell~2261 cluster and the magnification map of the cluster are shown in Figure~\ref{fig:a2261overview}. The coordinates of the other three areas were used to search for counter-images. All objects within 4$\arcsec$ of the predicted counter-image were considered. In the first region, within the blue circle in Figure~\ref{fig:a2261overview} the two objects \AbellOnFoSiSe{} ($1.7\arcsec$ angular distance to predicted counter-image position) and \AbellOnFoSiEi{} ($0.9\arcsec$ away in angular distance) were identified as possible counter-images. For the same reason as for the \AbellNiZeZeZe{} system, \AbellOnFoSiSe{} and \AbellOnFoSiEi{} are treated as a single object -- hereafter \AbellOnFoZeZeZe{} (see also Section~\ref{sec:abell14000}). In the second region \AbellOnThSiSi{}, $3.2\arcsec$ from the predicted coordinates, was identified as a plausible counter-image. For the last set of coordinates no suitable counter-image were found. Of these objects, \AbellNiOnZe{}, \AbellOnFoSiSe{} and \AbellOnThSiSi{} have previously been published in the high-redshift compilation by \citet{2014ApJ...792...76B}.

To rule out low-redshift old and/or dusty objects, data from four Spitzer filters was examined (Section~\ref{sec:observationaldata}). Visual inspection yielded no detections in these filters. This strengthens the case for the high-redshift solutions since low-redshift old and/or dusty objects would potentially be visible at the long wavelengths covered by Spitzer. Our optimizations also strongly prefer low or no dust solutions so including the Spitzer filters upper limits in our optimization do not change the results.

Extremely strong optical emission lines (such as [O\,\textsc{iii}] lines) from low-redshift objects can affect broadband photometry SED fitting and mimic a high-redshift Ly$\alpha$-line \citep{2015ApJ...801...12H}. But since there is continuum data for this candidate at longer wavelengths and it registers as a non-detection at wavelengths shorter than the Ly$\alpha$-break, it is likely not due to a strong non-Ly$\alpha$ emission line from a faint object. \citet{2015ApJ...801...12H} has conducted a search for extremely strong emission line galaxies of this type in CLASH and our candidate is not among those they identified, further strengthening the conclusion that our results are not due to strong optical emission line. Additionally, our spectroscopic observations, showing no prominent detection in the J-band (Section~\ref{sec:spectroscopicdata}), disfavor strong [O\,\textsc{ii}] or [O\,\textsc{iii}] from objects at $z \sim 2.1-4.8$.

\begin{figure*}
\begin{center}
\begin{tabular}{ | l | c | c | c | c | c | c | c | }
\hline
& F814W & F850LP & F105W & F110W & F125W & F140W & F160W \\
\hline
\begin{sideways}
{\scriptsize \AbellNiZeZeZe{}}
\end{sideways}
&
\begin{tikzpicture}
\node[anchor=south west,inner sep=0] (image) at (0,0) {\includegraphics[width=22mm, height=22mm]{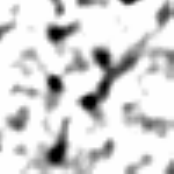}};
\draw[color=red!60, very thick](\TnNiUpx,\TnNiUpy) circle (\TnRad);
\draw[color=red!60, very thick](\TnNiLox,\TnNiLoy) circle (\TnRad);
\end{tikzpicture}
& 
\begin{tikzpicture}
\node[anchor=south west,inner sep=0] (image) at (0,0) {\includegraphics[width=22mm, height=22mm]{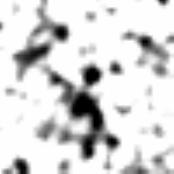}};
\draw[color=red!60, very thick](\TnNiUpx,\TnNiUpy) circle (\TnRad);
\draw[color=red!60, very thick](\TnNiLox,\TnNiLoy) circle (\TnRad);
\end{tikzpicture}
& 
\begin{tikzpicture}
\node[anchor=south west,inner sep=0] (image) at (0,0) {\includegraphics[width=22mm, height=22mm]{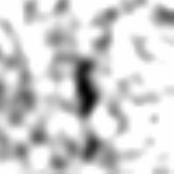}};
\draw[color=red!60, very thick](\TnNiUpx,\TnNiUpy) circle (\TnRad);
\draw[color=red!60, very thick](\TnNiLox,\TnNiLoy) circle (\TnRad);
\end{tikzpicture}
& 
\begin{tikzpicture}
\node[anchor=south west,inner sep=0] (image) at (0,0) {\includegraphics[width=22mm, height=22mm]{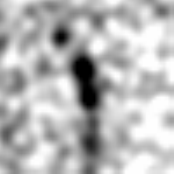}};
\draw[color=red!60, very thick](\TnNiUpx,\TnNiUpy) circle (\TnRad);
\draw[color=red!60, very thick](\TnNiLox,\TnNiLoy) circle (\TnRad);
\end{tikzpicture}
& 
\begin{tikzpicture}
\node[anchor=south west,inner sep=0] (image) at (0,0) {\includegraphics[width=22mm, height=22mm]{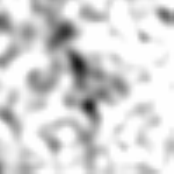}};
\draw[color=red!60, very thick](\TnNiUpx,\TnNiUpy) circle (\TnRad);
\draw[color=red!60, very thick](\TnNiLox,\TnNiLoy) circle (\TnRad);
\end{tikzpicture}
& 
\begin{tikzpicture}
\node[anchor=south west,inner sep=0] (image) at (0,0) {\includegraphics[width=22mm, height=22mm]{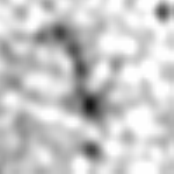}};
\draw[color=red!60, very thick](\TnNiUpx,\TnNiUpy) circle (\TnRad);
\draw[color=red!60, very thick](\TnNiLox,\TnNiLoy) circle (\TnRad);
\end{tikzpicture}
& 
\begin{tikzpicture}
\node[anchor=south west,inner sep=0] (image) at (0,0) {\includegraphics[width=22mm, height=22mm]{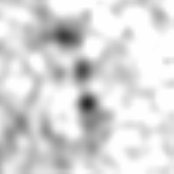}};
\draw[color=red!60, very thick](\TnNiUpx,\TnNiUpy) circle (\TnRad);
\draw[color=red!60, very thick](\TnNiLox,\TnNiLoy) circle (\TnRad);
\end{tikzpicture}
\\
\hline
\begin{sideways}
{\scriptsize \AbellOnFoZeZeZe{}}
\end{sideways}
&
\begin{tikzpicture}
\node[anchor=south west,inner sep=0] (image) at (0,0) {\includegraphics[width=22mm, height=22mm]{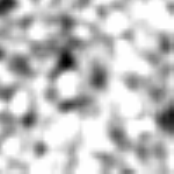}};
\draw[color=red!60, very thick](\TnFoUpx,\TnFoUpy) circle (\TnRad);
\draw[color=red!60, very thick](\TnFoLox,\TnFoLoy) circle (\TnRad);
\end{tikzpicture}
& 
\begin{tikzpicture}
\node[anchor=south west,inner sep=0] (image) at (0,0) {\includegraphics[width=22mm, height=22mm]{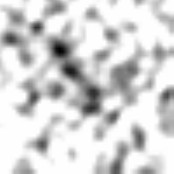}};
\draw[color=red!60, very thick](\TnFoUpx,\TnFoUpy) circle (\TnRad);
\draw[color=red!60, very thick](\TnFoLox,\TnFoLoy) circle (\TnRad);
\end{tikzpicture}
& 
\begin{tikzpicture}
\node[anchor=south west,inner sep=0] (image) at (0,0) {\includegraphics[width=22mm, height=22mm]{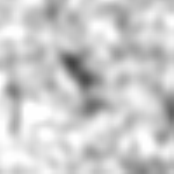}};
\draw[color=red!60, very thick](\TnFoUpx,\TnFoUpy) circle (\TnRad);
\draw[color=red!60, very thick](\TnFoLox,\TnFoLoy) circle (\TnRad);
\end{tikzpicture}
& 
\begin{tikzpicture}
\node[anchor=south west,inner sep=0] (image) at (0,0) {\includegraphics[width=22mm, height=22mm]{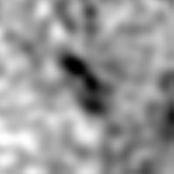}};
\draw[color=red!60, very thick](\TnFoUpx,\TnFoUpy) circle (\TnRad);
\draw[color=red!60, very thick](\TnFoLox,\TnFoLoy) circle (\TnRad);
\end{tikzpicture}
& 
\begin{tikzpicture}
\node[anchor=south west,inner sep=0] (image) at (0,0) {\includegraphics[width=22mm, height=22mm]{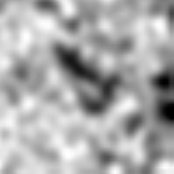}};
\draw[color=red!60, very thick](\TnFoUpx,\TnFoUpy) circle (\TnRad);
\draw[color=red!60, very thick](\TnFoLox,\TnFoLoy) circle (\TnRad);
\end{tikzpicture}
& 
\begin{tikzpicture}
\node[anchor=south west,inner sep=0] (image) at (0,0) {\includegraphics[width=22mm, height=22mm]{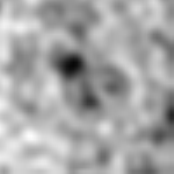}};
\draw[color=red!60, very thick](\TnFoUpx,\TnFoUpy) circle (\TnRad);
\draw[color=red!60, very thick](\TnFoLox,\TnFoLoy) circle (\TnRad);
\end{tikzpicture}
& 
\begin{tikzpicture}
\node[anchor=south west,inner sep=0] (image) at (0,0) {\includegraphics[width=22mm, height=22mm]{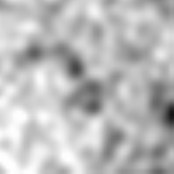}};
\draw[color=red!60, very thick](\TnFoUpx,\TnFoUpy) circle (\TnRad);
\draw[color=red!60, very thick](\TnFoLox,\TnFoLoy) circle (\TnRad);
\end{tikzpicture}
\\
\hline
\begin{sideways}
{\scriptsize \AbellOnThSiSi{}}
\end{sideways}
&
\begin{tikzpicture}
\node[anchor=south west,inner sep=0] (image) at (0,0) {\includegraphics[width=22mm, height=22mm]{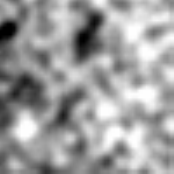}};
\draw[color=red!60, very thick](\TnThUpx,\TnThUpy) circle (\TnRad);
\draw[color=red!60, very thick](\TnThLox,\TnThLoy) circle (\TnRad);
\end{tikzpicture}
& 
\begin{tikzpicture}
\node[anchor=south west,inner sep=0] (image) at (0,0) {\includegraphics[width=22mm, height=22mm]{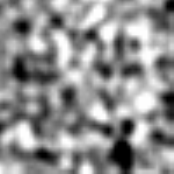}};
\draw[color=red!60, very thick](\TnThUpx,\TnThUpy) circle (\TnRad);
\draw[color=red!60, very thick](\TnThLox,\TnThLoy) circle (\TnRad);
\end{tikzpicture}
& 
\begin{tikzpicture}
\node[anchor=south west,inner sep=0] (image) at (0,0) {\includegraphics[width=22mm, height=22mm]{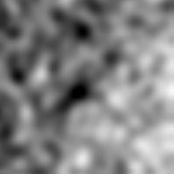}};
\draw[color=red!60, very thick](\TnThUpx,\TnThUpy) circle (\TnRad);
\draw[color=red!60, very thick](\TnThLox,\TnThLoy) circle (\TnRad);
\end{tikzpicture}
& 
\begin{tikzpicture}
\node[anchor=south west,inner sep=0] (image) at (0,0) {\includegraphics[width=22mm, height=22mm]{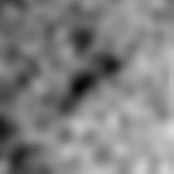}};
\draw[color=red!60, very thick](\TnThUpx,\TnThUpy) circle (\TnRad);
\draw[color=red!60, very thick](\TnThLox,\TnThLoy) circle (\TnRad);
\end{tikzpicture}
& 
\begin{tikzpicture}
\node[anchor=south west,inner sep=0] (image) at (0,0) {\includegraphics[width=22mm, height=22mm]{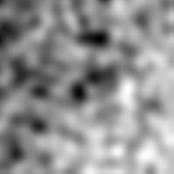}};
\draw[color=red!60, very thick](\TnThUpx,\TnThUpy) circle (\TnRad);
\draw[color=red!60, very thick](\TnThLox,\TnThLoy) circle (\TnRad);
\end{tikzpicture}
& 
\begin{tikzpicture}
\node[anchor=south west,inner sep=0] (image) at (0,0) {\includegraphics[width=22mm, height=22mm]{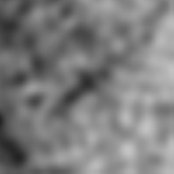}};
\draw[color=red!60, very thick](\TnThUpx,\TnThUpy) circle (\TnRad);
\draw[color=red!60, very thick](\TnThLox,\TnThLoy) circle (\TnRad);
\end{tikzpicture}
& 
\begin{tikzpicture}
\node[anchor=south west,inner sep=0] (image) at (0,0) {\includegraphics[width=22mm, height=22mm]{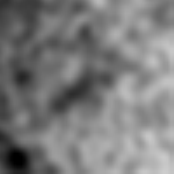}};
\draw[color=red!60, very thick](\TnThUpx,\TnThUpy) circle (\TnRad);
\draw[color=red!60, very thick](\TnThLox,\TnThLoy) circle (\TnRad);
\end{tikzpicture}
\\
\hline
\end{tabular}
\caption{
3.3$\arcsec$ $\times$ 3.3$\arcsec$ thumbnail images of the counter-images. Colors are inverted. Each column represents one of the seven filters with wavelengths relevant for $z>6$ objects. The Ly$\alpha$-break is evident between the filters F814W and F850LP for \AbellNiZeZeZe{} and \AbellOnFoZeZeZe{} but not as clearly for \AbellOnThSiSi{}. For all three objects hints of flux can be discerned in F814W which could be spurious detections or a smaller fraction of the spectrum visible in the filter. There is some substructure for \AbellNiZeZeZe{} and \AbellOnFoZeZeZe{} (both were identified originally as two objects each) in the form of two detected objects each in F125W, F140W and F160W marked by the red circles. In the shorter wavelength filters the objects look more like extended arcs. For \AbellOnThSiSi{} the substructure is not as clearly identifiable but can be seen in F105W and to some extent in F850LP and it resembles an arc in the other IR filters. The uppermost third object seen in \AbellNiZeZeZe{} was identified as a low redshift interloper, a conclusion strengthened by the fact that only two objects is observed in \AbellOnFoZeZeZe{}. In the filters F105W, F110W and F140W for \AbellNiZeZeZe{} an object can also be seen in the lower part of the thumbnails. However, it is significantly fainter than the other objects and its data and model fits do not correspond well to the other objects, so we discard it as part of \AbellNiZeZeZe{}. Comparing the extended arcs/substructure of the objects to Figure~\ref{fig:a2261overview}, they can be seen to be parallel to the critical curve. This implies that the gravitational lensing could be the source of the extension/increase the separation of the substructure.
}
\label{fig:RF1}
\end{center}
\end{figure*}

In addition to examine each one individually and compare them, we analyze the images as one object.  We sum the fluxes of all three of them in each filter, henceforth called \AbellStacked{}. Table~\ref{tab:RT1} lists the coordinates of the objects (including the stacked object) and their AB-magnitudes in the seven longest-wavelength CLASH filters. It also contains the redshift, metallicity, the $\chi^2$ value of the fit and the model implied rest-frame EW(Ly$\alpha$) corrected for ISM/IGM absorption using $f_{\mathrm{Ly}\alpha}$. The magnification estimated with the gravitational lens model is also included (Section~\ref{sec:counterimagecomparison}). Figure~\ref{fig:RF1} shows thumbnail images of the objects.

\begin{table*}
\caption{
Coordinates, photometric AB-magnitudes, redshift, metallicity, reduced $\chi^2$s of the best-fitting model (as a familiar measure of the quality of fit) and model implied rest-frame equivalent width for Ly$\alpha$ corrected for ISM/IGM absorption using $f_{\mathrm{Ly}\alpha}$. $\mu_{z=6.3}$ is the magnification corresponding to the \AbellStacked{}s redshift of 6.3.
}
\label{tab:RT1}
\begin{center}
\begin{tabular}{lcccc}
\hline
\\
Object & \AbellNiZeZeZe{} & \AbellOnFoZeZeZe{} & \AbellOnThSiSi{} & \AbellStacked{} \\
\hline
\\
Right Ascension & $ 17^{\mathrm{h}}22^{\mathrm{m}}25.13^{\mathrm{s}} $ & $ 17^{\mathrm{h}}22^{\mathrm{m}}28.48^{\mathrm{s}} $ & $ 17^{\mathrm{h}}22^{\mathrm{m}}25.85^{\mathrm{s}} $ & - \\
Declination & $ 32^{\circ}08\arcmin08~3\arcsec$ & $ 32^{\circ}07\arcmin33~3\arcsec$ & $ 32^{\circ}07\arcmin39~8\arcsec$ & - \\
F814W & $ 27.92 \pm 0.43 $ & $ 26.68 \pm 0.20 $ & $ 27.75 \pm 0.37 $ & $ 26.11 \pm 0.18 $ \\
F850LP & $ 25.93 \pm 0.17 $ & $ 24.78 \pm 0.12 $ & $ 27.63 \pm 0.68 $ & $ 24.40 \pm 0.10 $ \\
F105W & $ 26.17 \pm 0.12 $ & $ 25.71 \pm 0.10 $ & $ 26.2 \pm 0.12 $ & $ 24.80 \pm 0.06 $ \\
F110W & $ 26.04 \pm 0.08 $ & $ 25.89 \pm 0.08 $ & $ 25.99 \pm 0.08 $ & $ 24.78 \pm 0.05 $ \\
F125W & $ 26.27 \pm 0.14 $ & $ 25.83 \pm 0.11 $ & $ 26.58 \pm 0.19 $ & $ 24.99 \pm 0.08 $ \\
F140W & $ 26.28 \pm 0.11 $ & $ 25.87 \pm 0.10 $ & $ 26.34 \pm 0.13 $ & $ 24.95 \pm 0.07 $ \\
F160W & $ 26.37 \pm 0.13 $ & $ 25.99 \pm 0.11 $ & $ 26.29 \pm 0.13 $ & $ 25.01 \pm 0.07 $ \\
$z$ & $ 6.4 $ & $ 6.2 $ & $ 6.8 $ & $ 6.3 $ \\
Z & $ 0 $ & $ 0 $ & 0.02~$\mathrm{Z}_{\odot}$ & 0 \\
$\chi^2$ & $ 0.85 $ & $ 1.74 $ & $ 2.76 $ & 1.37 \\
EW(Ly$\alpha$) & $ 210~\AA{} $ & $ 330~\AA{} $ & $ 200~\AA{} $  & $ 160~\AA{} $\\
$\mu_{z=6.3}$ & $ \AbellNiZeZeZeMagnification{} $ & $ \AbellOnFoZeZeZeMagnification{} $ & $ \AbellOnThSiSiMagnification{} $ & - \\
\\
\hline
\end{tabular}
\end{center}
\end{table*}

\begin{figure*}
     \begin{center}
			\includegraphics[width=135mm]{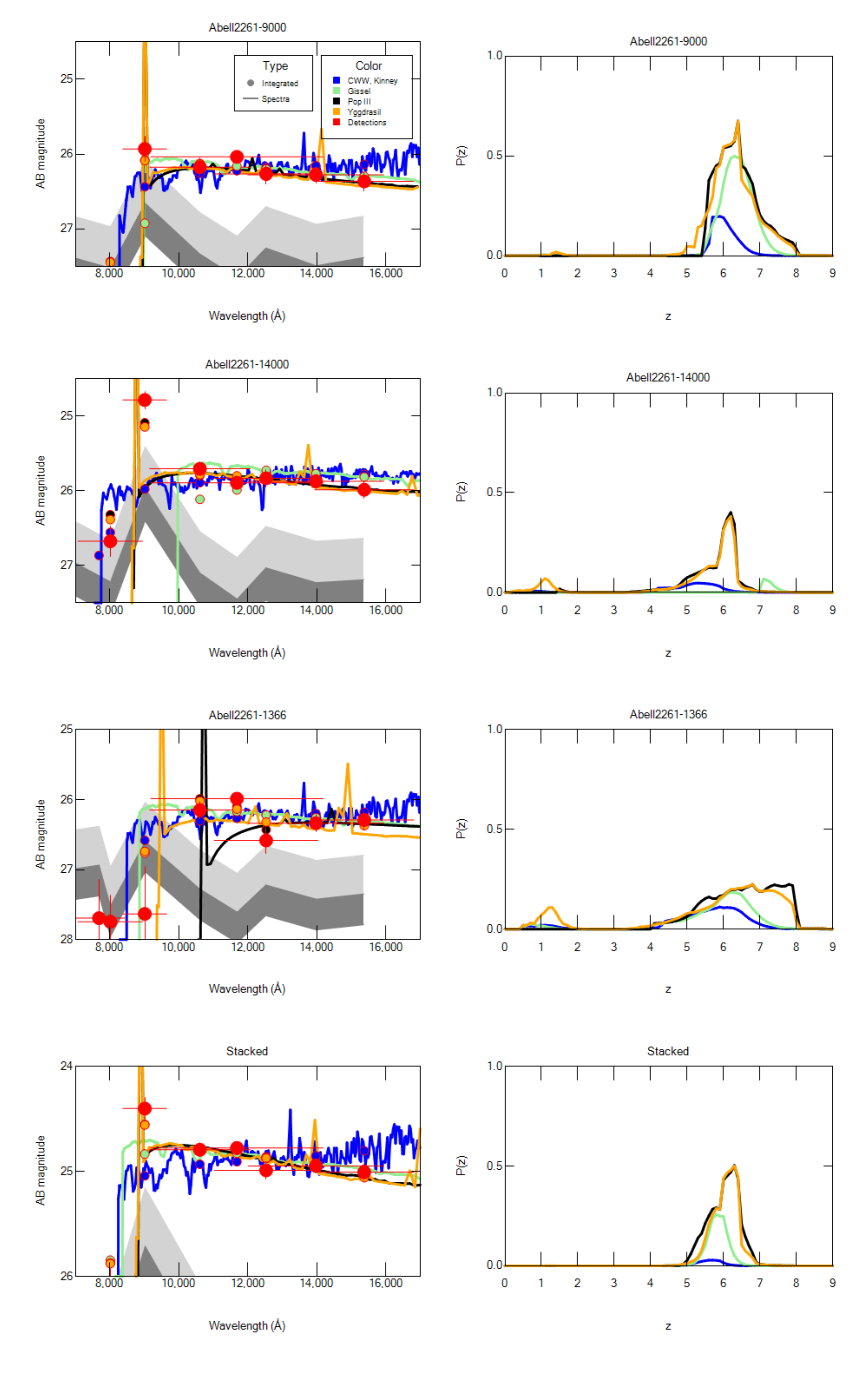}
      \caption{ 
Left column: magnitudes and synthetic spectra of best fitted model for each grid. Limiting magnitudes are shown in dark gray for $2\sigma - 3\sigma$ detections and light gray for $3\sigma - 5\sigma$ detections. Right column: P($z$) plots comparing the four grids of models. The model with the best quality of fit for each $z$ is shown. All three images and the stacked image exhibit a Ly$\alpha$-break in the best fitted models in the range $\lambda \sim 8,000-10,000$~\AA{}. The Pop~III grid for \AbellOnThSiSi{} has a higher Ly$\alpha$-break and hence an even higher photometric redshift but the Yggdrasil grid has slightly better quality of fit. The large uncertainties of the redshift estimate for  \AbellOnThSiSi{} can be seen in the right column. The other two images and the stacked image have significantly smaller uncertainties.
}
      \label{fig:RF2}
   \end{center}
\end{figure*}

The P($z$) and P($f_{\mathrm{Ly}\alpha}$) quantities, i.e. the photometric-redshift distribution and the photometric-$f_{\mathrm{Ly}\alpha}$ distribution, are calculated using the cross-validation technique described in \citet{2015ApJ...804...13R}. For each $z$ or $f_{\mathrm{Ly}\alpha}$, the model (for the considered grid) with the highest cross-validation value (i.e. quality of fit) is selected. A good quality of fit for a certain grid thus means there is at least one model in the grid that fits the observations well. The cross-validation value is normalized to the interval [0,~1] (where 1 implies that each data point is perfectly reproduced by the model), and is used as a proxy for a probability distribution.

In Figure~\ref{fig:RF2}, each row corresponds to one of the three objects \AbellNiZeZeZe{}, \AbellOnFoZeZeZe{}, and \AbellOnThSiSi{} as well as the stacked version, \AbellStacked{}. The observational data and spectrum of the best fitted model for each grid are listed in the first column while the second column shows the P($z$) from different model grids.

\subsection{\AbellNiOnZe{} and \AbellNiOnOn{}}
\label{sec:abell9000}

As mentioned above, \AbellNiOnZe{} was originally identified as a Pop~III galaxy candidate. Another object, \AbellNiOnOn{}, adjacent to and with similar properties as \AbellNiOnZe{} was later identified and we refer to it as substructure of the same galaxy, or a member of an interacting galaxy pair. \AbellNiOnZe{} and \AbellNiOnOn{} will be hereafter treated as substructure of a single object, \AbellNiZeZeZe{}. To analyze \AbellNiOnZe{}/\AbellNiOnOn{} as one object the fluxes in each filter were added and new magnitudes were calculated. This method might differ in the flux an independent measurement would pick up since the aperture would differ. \AbellNiZeZeZe{} has 5$\sigma$ (10$\sigma$ in F110W) detections in F850LP and all IR filters. 

\subsection{\AbellOnFoSiSe{} and \AbellOnFoSiEi{}}
\label{sec:abell14000}

In the area predicted to contain a counter-image marked by blue in Figure~\ref{fig:a2261overview}, the official catalogs contained 8 objects. Of these all but two either had redshifts that were too low or were too faint (the detections were all below 5$\sigma$), and none had morphologies corresponding to \AbellNiOnZe{}/\AbellNiOnOn{}. The remaining two, \AbellOnFoSiSe{} and \AbellOnFoSiEi{}, are close enough to the predicted location to be the counter-image of \AbellNiZeZeZe{}, including the binary substructure. The images are $1.7\arcsec$ and $0.9\arcsec$ for \AbellOnFoSiSe{} and \AbellOnFoSiEi{}, respectively, from the predicted location.

As with \AbellNiZeZeZe{} the flux of \AbellOnFoSiSe{} and \AbellOnFoSiEi{} are added to represent one object, \AbellOnFoZeZeZe{}. Like \AbellNiZeZeZe{}, \AbellOnFoZeZeZe{} has 5$\sigma$ detections in F850LP as well as all IR filters, even 10$\sigma$ in F105W, F110W and F140W.

\subsection{\AbellOnThSiSi{}}

In the official catalogs seven objects exist within $4.0\arcsec$, encompassed by the purple circle in Figure~\ref{fig:a2261overview}, of the counter-image's predicted position. One of them is a star covering a large fraction of the area where the counter-image is predicted. One is another, small star, and four are either too faint or do not have matching morphology, photometric redshift and/or colors. The last object, \AbellOnThSiSi{}, found $3.2\arcsec$ from the position of the predicted counter-image, does correspond sufficiently well to \AbellNiZeZeZe{} and \AbellOnFoZeZeZe{} to be reliably considered as a counter-image.

\AbellOnThSiSi{} appears as an extended arc parallel to the critical curve (compare Figure~\ref{fig:RF1} to Figure~\ref{fig:a2261overview}) in the IR filters except F105W (and to some extent in F850LP) where substructure can be discerned. \AbellOnThSiSi{} has 5$\sigma$ (10$\sigma$ in F110W) detections in all IR filters. As opposed to \AbellNiZeZeZe{} and \AbellOnFoZeZeZe{} \AbellOnThSiSi{} has a non-detection in F850LP, the difference is discussed in Section~\ref{sec:counterimagecomparison}. This non-detection means the SED-fitting uses five filters instead of six and the image being slightly fainter is the plausible reason for its higher photometric redshift of 6.8 and generally lower quality of fit compared to \AbellNiZeZeZe{} and \AbellOnFoZeZeZe{}. Its higher photometric redshift does not make it incompatible as counter-image since its quality of fit is almost as good at the photometric redshifts corresponding to \AbellNiZeZeZe{} and \AbellOnFoZeZeZe{}.

The most plausible alternative explanation to \AbellOnThSiSi{} being a counter-image would be that the star (obstructs a circular area $r \sim 1\arcsec$) hides the counter-image.

\subsection{The third predicted counter-image}

The gray circle in Figure~\ref{fig:a2261overview} marks the third predicted counter-image. The official catalogs contain 9 objects within $4.0\arcsec$ of the predicted counter-image. Four of the objects are so faint ($\mathrm{S/N} < 5$ in all filters) as to be flagged as spurious detections in the official catalogs. Three objects are stars and the last two objects have too low photometric redshift. Since the predicted magnifications, $\mu \sim 2.5$, in the region are lower than for the other regions the counter-image is expected to be fainter. The magnitudes of the fainter objects, $m_{\mathrm{AB}} \sim 28.5$, are barely consistent with those of the other counter-images when considering the magnification including errors.

There are two possibilities for the counter-image: either it is one of the fainter objects and the magnification is significantly different from the calculated one (but still barely within the error bars) or it is behind one of the foreground stars. Since their morphology does not correspond to that of either \AbellNiZeZeZe{}, \AbellOnFoZeZeZe{} or \AbellOnThSiSi{} (as there are no two faint objects close) we will not consider the fainter objects further.

\subsection{Counter-image comparison}
\label{sec:counterimagecomparison}

When considering different images as counter-images of the same object, the images have to be consistent with regard to redshift as well as morphology and fluxes (including magnification).

As a first step, we have added the fluxes from the three counter-images and fit the resulting magnitudes to models, what we call \AbellStacked{}. One drawback of this approach is that the fitting might be dominated by one image, especially if it is significantly brighter. This means the fit might be good for only one image. However, our three counter-images have roughly similar magnification and magnitudes. We find a best fit at $z=6.3$ with good quality fits extending from $z_{\mathrm{L}} \approx 5.6$ to $z_{\mathrm{U}} \approx 6.5$ and EW(Ly$\alpha$) $\approx 160$~\AA{}.

\begin{figure}
  \begin{center}
	\includegraphics[width=80mm]{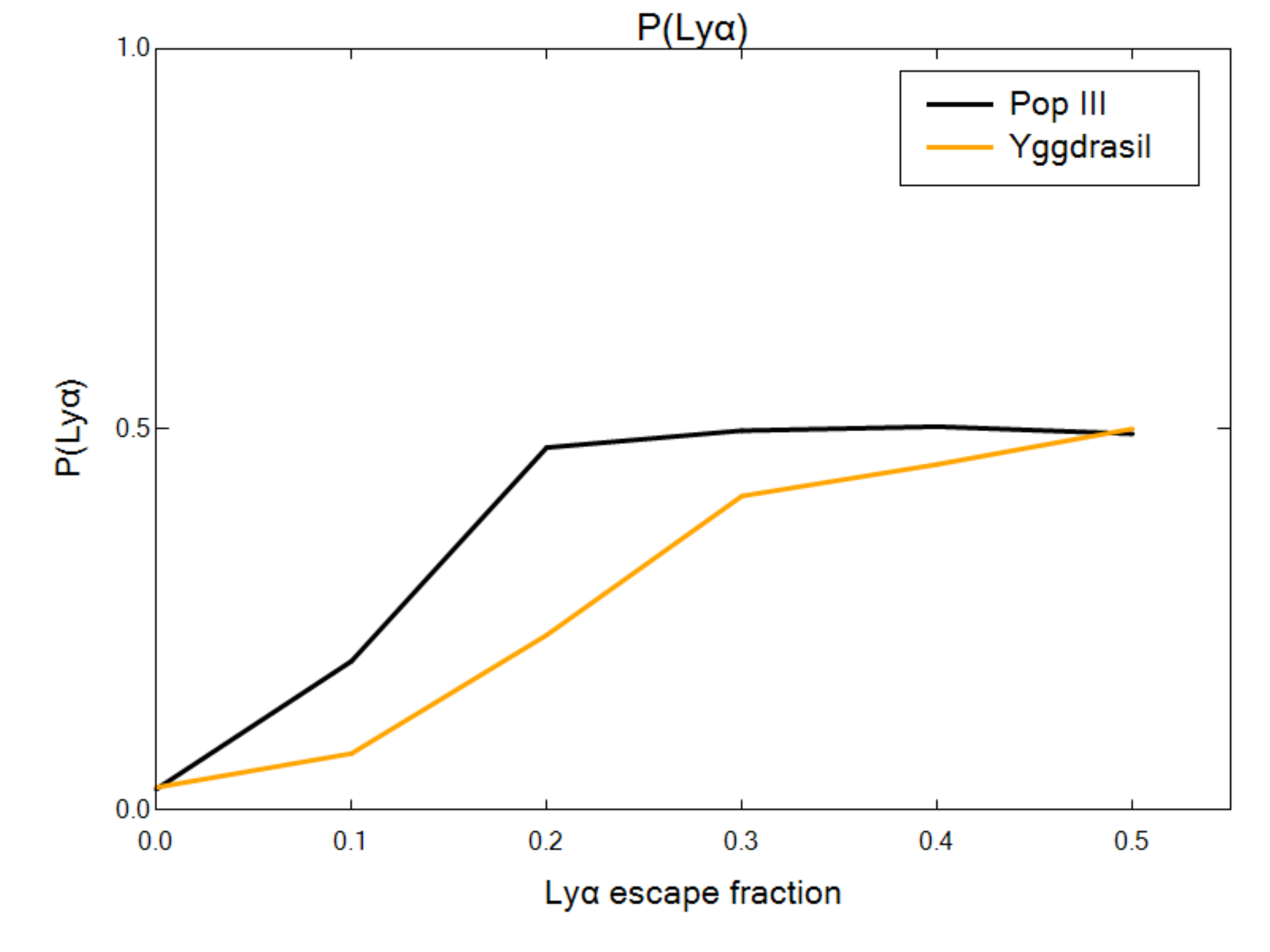}
  \end{center}
  \caption{
The quality of fit as a function of the Ly$\alpha$ escape fraction $f_{\mathrm{Ly}\alpha}$ at $z = 6.3$. As can be seen, it has a strong dependence on $f_{\mathrm{Ly}\alpha}$. The Pop~III allow a lower $f_{\mathrm{Ly}\alpha}$ with high quality of fit. The Yggdrasil $\mathrm{Z} > 0$ model grid is more sensitive to $f_{\mathrm{Ly}\alpha}$ and since galaxies with metals are more likely to contain dust the  $f_{\mathrm{Ly}\alpha}$ might be low. This could mean there is a, admittedly weak, case for it to be metal-free or at least having a very low metallicity.
}
  \label{fig:convolved}
\end{figure}

For \AbellStacked{} the most plausible redshifts are 6.3 for both the Pop~III grid and the Yggdrasil $\mathrm{Z}>0$ grid. The Gissel and CWW, Kinney grids are not considered further because their P($z$) are significantly lower compared to the Yggdrasil grids. Figure~\ref{fig:convolved} contains the quality of fits dependence on $f_{\mathrm{Ly}\alpha}$ for \AbellStacked{} at $z=6.3$. It is evident that both Yggdrasil grids have a strong dependence on $f_{\mathrm{Ly}\alpha}$. The model with best fit has EW(Ly$\alpha$)~$\approx$~160~\AA{}. Pop~III galaxy models and Population~II/I galaxy models have essentially equally good fits. Hence, there are alternatives to the Pop~III interpretation of the object.

\begin{figure*}
     \begin{center}
     \begin{tabular}{ | c | c | c | }
  {\Large \AbellNiZeZeZe{}}   & {\Large \AbellOnFoZeZeZe{}} & {\Large \AbellOnThSiSi{}} \\
\begin{tikzpicture}
\node[anchor=south west,inner sep=0] (image) at (0,0) {\includegraphics[width=45mm, height=45mm]{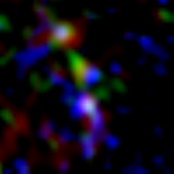}};
\draw[color=red!60, very thick](2.1,2.6) circle (0.4);
\draw[color=red!60, very thick](2.2,1.7) circle (0.4);
\end{tikzpicture}
     &
\begin{tikzpicture}
\node[anchor=south west,inner sep=0] (image) at (0,0) {\includegraphics[width=45mm, height=45mm]{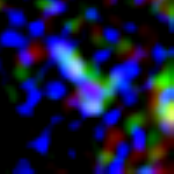}};
\draw[color=red!60, very thick](1.8,2.65) circle (0.4);
\draw[color=red!60, very thick](2.3,1.7) circle (0.4);
\end{tikzpicture}
&
\begin{tikzpicture}
\node[anchor=south west,inner sep=0] (image) at (0,0) {\includegraphics[width=45mm, height=45mm]{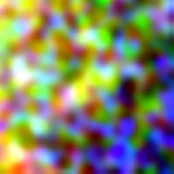}};
\draw[color=red!60, very thick](2.8,2.7) circle (0.4);
\draw[color=red!60, very thick](1.9,1.9) circle (0.4);
\end{tikzpicture}
      \\
\multicolumn{3}{c}{\includegraphics[width=140mm]{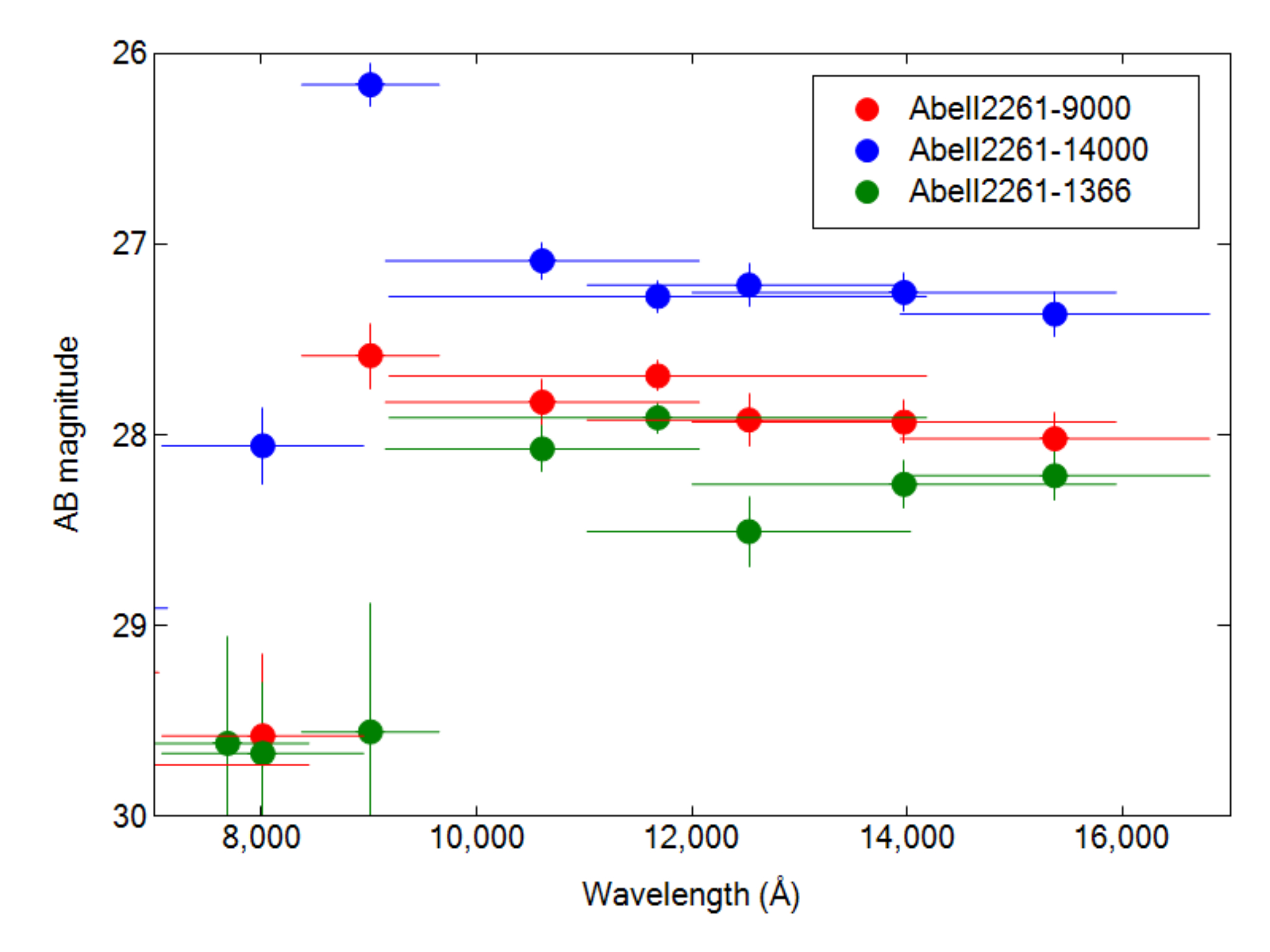}}\\
      \end{tabular}
      \caption{
The first row contain multi-color (F850LP is represented by blue, F125W is represented by green, and F160W is represented by red) thumbnails of the three counter-images. The second row contains a graph with the observed AB-magnitudes for the three images delensed with our magnification estimates. \AbellOnFoZeZeZe{} is extremely bright in F850LP which is reflected in the thumbnail being very blue. The thumbnail image's blue part is also large compared to \AbellNiZeZeZe{} and \AbellOnThSiSi{} and connects the substructure. This could indicate that the Ly$\alpha$ halo is above the detection limit and thus observed in this counter-image. \AbellOnThSiSi{} is polluted by a nearby star which makes the thumbnail background bright. This could render the extended radiation indistinguishable from the background and thus unobserved, explaining the faintness of \AbellOnThSiSi{} in F850LP. The general form of the observations in the other filters, i.e. fainter with wavelength in IR, coincides for the different images. However, \AbellOnFoZeZeZe{} is $\sim 1$~AB-magnitudes, brighter than the other counter-images, corresponding to a factor 2.5. For the magnitudes the errors in magnifications is systematic and large enough to bridge the difference in magnitude between the different images.
}
      \label{fig:magnificationcomparison}
      \end{center}
\end{figure*}
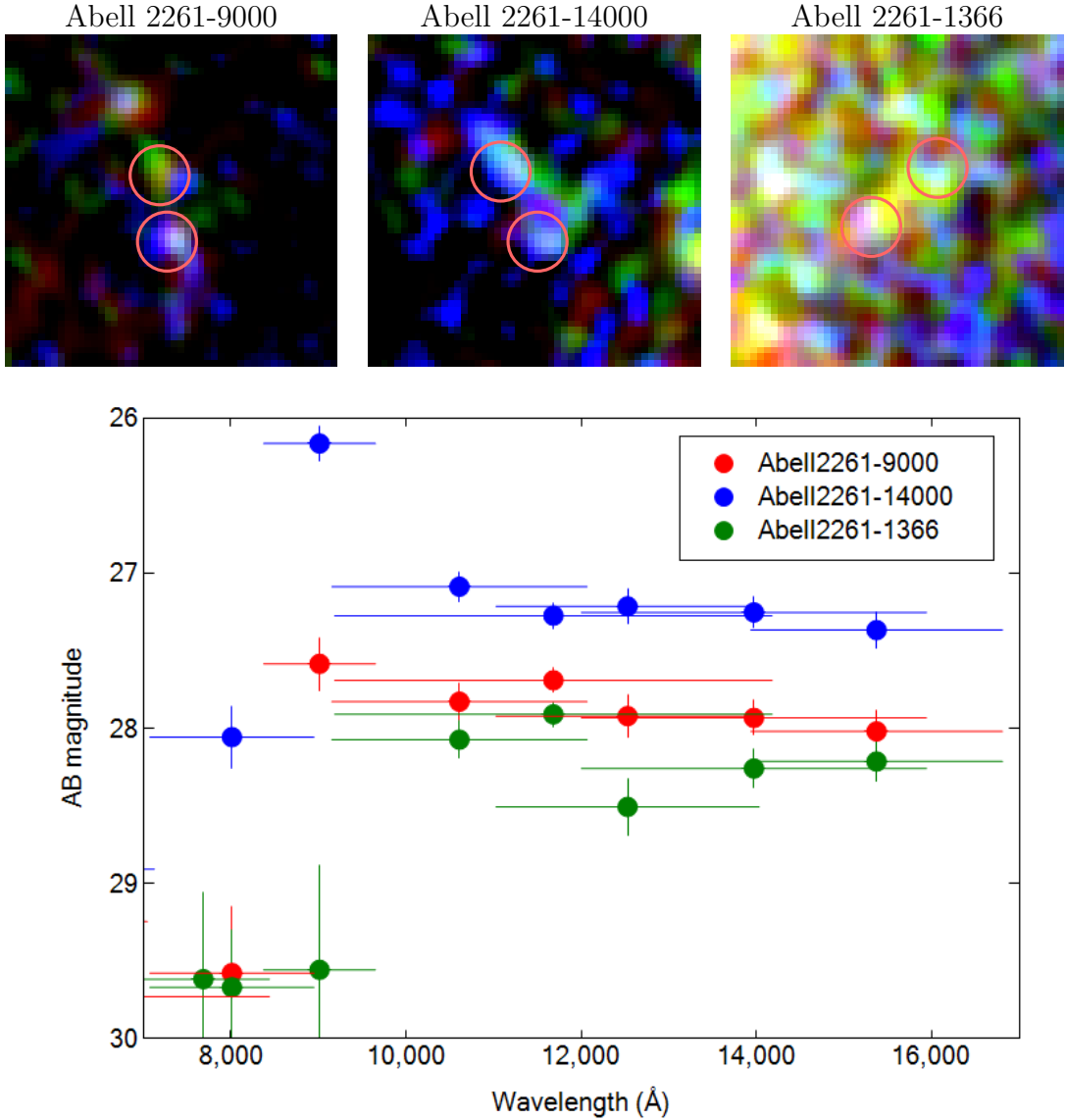

Figure~\ref{fig:magnificationcomparison} compares the magnitudes of the three counter-images after de-lensing them using our magnification estimates. We see that \AbellOnFoZeZeZe{} is approximately 1.0 magnitudes (a factor 2.5) brighter than \AbellNiZeZeZe{} and \AbellOnThSiSi{} after de-lensing. But the errors in the magnification estimates, which are systematic across the filters, are large and the discrepancies are within the errors. The observations in F850LP, however, has a too large difference to be explained by magnification errors. Also, when comparing the colors in Figure~\ref{fig:magnificationcomparison} the observations in the IR filters are consistent with an SED with declining slope while only the F850LP observation for \AbellNiZeZeZe{} is, \AbellOnFoZeZeZe{}s F850LP magnitude is much brighter while \AbellOnThSiSi{}s F850LP magnitude is much fainter. This poses a challenge for the interpretation but it could possibly be explained by an extended Ly$\alpha$ halo around the galaxy. The sensitivity of the detector could be such that the halo is only detected if the object is bright enough, hence contributing a disproportional amount of flux for brighter objects. The flux is consistent with this since it is significantly brighter in the brightest image \AbellOnFoZeZeZe{} compared to \AbellNiZeZeZe{} and \AbellOnThSiSi{}. The reason \AbellOnThSiSi{} differs from \AbellNiZeZeZe{} could be that \AbellOnThSiSi{} is contaminated by light from the nearby star, rendering the halo indistinguishable from the illumination provided by the star.

\subsection{Interacting galaxies?}

To constitute a merger or a virialized galaxy the substructure must be within the virialized radius of the DM halo. Hence, if the distance between the two clumps is larger than twice the virialized radius the two clumps are not part of the same entity nor in the process of becoming one (even though a future collision might result in a merger). The lowest mass estimate of our models is $\sim10^6~\mathrm{M}_{\odot}$ in stars. If 15\% of the DM halo is baryonic and assuming all baryons are in stars (in reality only a small fraction will be in stars which would mean an even more massive DM halo) this correspond to a $7 \times 10^6~\mathrm{M}_{\odot}$ DM halo. The virialized mass we use is the M$_{200}$ mass approximation, the mass that resides within the radius at which the density is 200 times the critical density which also correspond to a virialized radius R$_{200}$. Approximating our DM halo mass with M$_{200}$ we arrive at R$_{200} = 800$~pc. Since our assumptions minimize the DM halo mass this is the minimum virialized radius according to our mass estimates.

The angular distances between the centers of substructure in the three images (0.60$\arcsec$ for \AbellNiZeZeZe{}, 0.87$\arcsec$ for \AbellOnFoZeZeZe{} and 1.2$\arcsec$ for \AbellOnThSiSi{}) which corresponds to delensed distances expressed as radii of $\gtrsim 370$~pc at $z = 6.3$. Since the distance between the substructures along the sightline is unknown this should be taken as a lower limit. Comparing this to the minimum virialized radius of R$_{200} = 800$~pc we conclude that the system could be close enough to be virialized and therefore substructure in a galaxy or an interaction/merger. 

The substructures proximity could indicate a galaxy merger \citep{2004ApJ...617L...9L} but could also be a chance projection of two high-redshift objects in which the distance between them along the sightline is too large to be a merger. However, given the similarities in morphology, redshift estimate and colors between the three images it seems implausible they could be chance projections along the line of sight. The AB-magnitude increase/decrease due to differing distance would have to be compensated by correspondingly higher/lower luminosity. The angular size of the substructures are also roughly equal which would have to be a coincidence if the substructure were chance projections. Because our current data cannot constrain collision angle or speed, composition (which we have assumed to be the same in our combined fitting) or relative mass, further investigation is required to determine if the lensed object is a merger.

\section{Summary and conclusion}
\label{sec:summaryconclusion}

We have identified a likely, multiply-lensed LAE candidate in the CLASH survey. This object has three counter-image candidates identified, \AbellNiZeZeZe{}, \AbellOnFoZeZeZe{} and \AbellOnThSiSi{}, with magnifications $\mu \approx \AbellNiZeZeZeMagnification{}$, $\AbellOnFoZeZeZeMagnification{}$ and $\AbellOnThSiSiMagnification{}$, respectively. Another image is predicted but is not detected.
In the process of locating the counter-images we measured the first spectroscopic redshift ($z = 3.377$) of a multiply imaged galaxy behind Abell~2261, which was used to calibrate the lensing model. The LAE was initially considered to be a Pop~III galaxy candidate but competitive fits with non-zero metallicity models indicate other interpretations are equally likely.  By combining these fits using Yggdrasil models a redshift estimate of 6.3 was obtained for the galaxy. This redshift estimate is dependent on the assumption of strong Ly$\alpha$ emission.

In two of the counter-images, substructure appearing as two objects in the same orientation expected from the lens model in the filters with longest wavelength were found. The substructure could also be discerned in the third image even though it was not separable into two objects. They could be star-forming regions belonging to the same galaxy or two interacting galaxies. SED fitting of models to photometry implies extreme Ly$\alpha$ emission EW(Ly$\alpha$)~$\sim 160$~\AA{}. We also find that the angular distance spanning its substructures is not too large for them to be a single virialized structure or a merger between two small galaxies. The kinematics of the structures could not be constrained with current observations but might be with additional surveys, particularly when using the [C\,\textsc{ii}] line. Splitting in this line would indicate large relative velocities that may indicate a recent merger.

This galaxy joins an increasing group of Ly$\alpha$ emitters now being discovered at $z > 6$. Similar objects at higher redshifts may be the birthplaces of direct collapse black holes, the likely precursors of the most massive black holes at $z > 6$ \citep{2011Natur.474..616M,2015Natur.518..512W}. Others, as may be true of CR7, may harbor Pop III stars. CLASH, Frontier Fields and future surveys of cluster lenses may reveal these objects and yield clues to the origins of the first stars and supermassive black holes.

\section*{Acknowledgments}

CER acknowledges funding from the Swedish National Space Board and the Royal Swedish Academy of Sciences. CER, RSK, and DJW were supported by the European Research Council under the European Community's Seventh Framework Programme (FP7/2007 - 2013) via the ERC Advanced Grant ``STARLIGHT: Formation of the First Stars" (project number 339177). EZ acknowledges financial support from the Swedish National Space Board, the Swedish Research Council (project 2011-5349) and the Wenner-Gren Foundations. RSK furthermore is grateful for funding from the {\em Deutsche Forschungsgemeinschaft} DFG via SFB 881, 'The Milky Way System' (sub-projects B1, B2 and B8) and from SPP 1573 'Physics of the Interstellar Medium' (projects KL 1358/18.1, KL 1358/19.2). Support for AZ was provided by NASA through Hubble Fellowship grant \#HST-HF2-51334.001-A awarded by STScI. The Keck observations were conducted by AZ together with Richard Ellis and Sirio Belli, and we thank their respective role. J.G. acknowledges financial support from the Wenner-Gren Foundations.

Some of the data presented herein were obtained at the W.M. Keck Observatory, which is operated as a scientific partnership among the California Institute of Technology, the University of California and the National Aeronautics and Space Administration. The Observatory was made possible by the generous financial support of the W.M. Keck Foundation. AZ wishes to recognize and acknowledge the very significant cultural role and reverence that the summit of Mauna Kea has always had within the indigenous Hawaiian community.  We are most fortunate to have the opportunity to conduct observations from this mountain.




\bibliographystyle{mnras}
\bibliography{References}



\bsp	
\label{lastpage}
\end{document}